\documentclass[aps,prd,10pt,notitlepage,nofootinbib,longbibliography]{revtex4-1}

\usepackage{amsmath, graphicx, 
}

\def\<{\langle}
\def\>{\rangle}

\begin{document}

\title{AdS-Sliced Flavor Branes and Adding Flavor to the Janus Solution}

\author{Adam B. Clark}
 \email{aclark@muhlenberg.edu}
\affiliation{Muhlenberg College}
\author{Nathan Crossette}
\email{nc242671@muhlenberg.edu} \affiliation{Muhlenberg College}
\author{George M. Newman}
\email{gmnewman@wsuvancouver.edu}
\affiliation{Washington State University - Vancouver}
\author{Andrea Rommal}
\email{ar244930@muhlenberg.edu}
\affiliation{Muhlenberg College}
\date{\today}

\begin{abstract}
We implement D7 flavor branes in AdS-sliced coordinates on $AdS_5\times
S^5$ with the ansatz that the brane fluctuates only in the warped ($\mu$)
direction in this slicing, which is particularly appropriate for studying
the Janus solution. The natural field theory dual in this slicing is
$\mathcal{N}=4$ super Yang-Mills on two copies of $AdS_4$. Branes
extending from $\mu=\pm\pi/2$ can end at different locations, giving rise
to quarks with piecewise constant mass on each $AdS_4$ half-space, jumping
discontinuously between them. A second class of flavor brane solutions
exists in this coordinate system, dubbed ``continuous'' flavor branes,
that extend across the entire range of $\mu$.  We propose that the correct
dual interpretation of ``disconnected'' flavor brane in this ansatz is a
quark hypermultiplet with constant mass on one of the AdS$_4$ half-spaces
with totally reflecting boundary conditions at the boundary of AdS$_4$;
whereas the dual interpretation of a continuous flavor brane has totally
transparent boundary conditions. Numerical studies indicate that
AdS-sliced D7 flavor branes of both classes exhibit spontaneous chiral
symmetry breaking, as non-zero vev persists for solutions of the equation
of motion down to zero mass.  Continuous flavor branes in this ansatz
exhibit many other surprising behaviors:  their masses seem to be capped
at a modest value near $m=0.551$ in units of the inverse AdS radius, and
there may be a phase transition between continuous branes of different
configurations. We also numerically study quark states in Janus.  The
behavior of mass and vev is similar in Janus, including the existence of
chiral symmetry breaking at zero mass, though qualitative features of the
phase diagram change (sometimes significantly) as the Janus parameter
$c_0$ increases.
\end{abstract}

\maketitle

\section{Introduction}
Holography and the AdS/CFT correspondence have been exciting fields of study
since their discovery late nineties \cite{Maldacena,wittenholography,gkp}.  Matter fields can be added to the gravity side by
the well-known prescription of adding a small number, $N_f$, of ``flavor''
branes~\cite{kk}.  This breaks half the supersymmetries of the gauge theory.  The
Janus solution of type IIB supergravity is interesting because locally it is
very similar to ordinary AdS space, but the dilaton and 5-form vary in the
warped direction in a particular slicing.  This situation breaks
supersymmetry entirely yet remains stable.  The dual gauge theory is most
commonly viewed as an ``interface'' conformal field theory, or ICFT, that is
$\mathcal{N}=4$ super Yang-Mills theory with a gauge coupling that jumps
suddenly at the interface of a domain wall.  The jumping coupling is
somewhat analogous to filling half the universe with a dielectric medium with a planar interface. This analogy is imprecise in one very important way: the speed of light does not change when crossing the domain wall, as it would if the interface were truly that between two dielectric media.

We expect systems that break supersymmetry to generically also break chiral
symmetry as the formation of a condensate is no longer forbidden.  Janus
provides a novel mechanism for this to occur, since supersymmetry in the
dual gauge theory is broken by the jumping of the coupling constant.  Chiral
symmetry will be broken by the same unusual mechanism.  More surprisingly,
regular AdS space with flavor branes set up according to a Janus-like ansatz \emph{also}
appears to break chiral symmetry.

\section{Review of Janus Solution}
\subsection{Coordinate Systems}
We review a variety of coordinate systems for $AdS_{d+1}$, defined as a hyperboloid in $R^{2,d}$:
\begin{equation}
X_0^2 + X_{d+1}^2 - X_1^2 - \dots - X_d^2=1.
\end{equation}
We will always work with unit radius.  If necessary, the AdS curvature
radius can be restored  with dimensional analysis.
\subsubsection{Global coords}
Global coordinates cover the entirety of AdS space. They consist of the following parametrization:
\begin{equation}
X_0 = \frac{\cos \tau}{\cos \theta} , \ \ X_{d+1} = \frac{\sin \tau}{\cos \theta} ,
\ \ X_i = \tan\theta n_i, \ i= 1, \dots , d,
\end{equation}
where the $n_i$ are unit vectors on $R^{d}$.  The coordinate $\tau$ fills
the time-like role,  and $\theta$ is the warped direction and is bounded
$[0,\pi/2]$.  The metric in these coordinates is
\begin{equation}
ds^2_{AdS_{d+1}} = \frac{1}{\cos^2 \theta} \left( - d\tau^2 + d\theta^2 + \sin^2 \theta d\Omega^2_{d-1}
\right).
\end{equation}
\subsubsection{Poincar\'e patch coordinates}
Often the most convenient coordinates despite an ugly parametrization:
\begin{eqnarray}
& X_0 = \frac{1}{2} \left(z + (1+ \vec{x}^2 - t^2) \right), \ \ X_{d+1} = \frac{t}{z},
\ \ X_d = \frac{1}{2} \left(z - (1- \vec{x}^2 + t^2) \right), & \\
\nonumber & \ \ X_{i=1,\dots , d-1} = \frac{x_i}{z}. &
\end{eqnarray}
This gives us the metric:
\begin{equation}
ds^2_{AdS_{d+1}} = \frac{1}{z^2} \left( - dt^2 + d \vec{x}^2 + dz^2
\right).
\end{equation}
The timelike coordinate is of course $t$, $z$ is the warped coordinate
bounded $[0,\infty]$,  and $\vec{x}$ denotes $(x_1,\dots x_{d-1})$.  Patch
coordinates only cover half of the spacetime.  If one defines
$u=\frac{1}{z}$, Poincar\'e patch coordinates can be recast so that the
metric takes on the form:
\begin{equation}
ds^2_{AdS_{d+1}} = u^2 \left( - dt^2 + d \vec{x}^2 \right) + \frac{1}{u^2} du^2.
\end{equation}
When it is necessary to distinguish these two coordinate systems we will
refer to the  coordinates with $u$ as ``braneworld patch coordinates,'' as
this is the metric that shows up most naturally when taking the near horizon
limit of the spacetime created by a stack of concident D-branes.  The
coordinate $u$ plays the role of a radius or transverse distance from the
branes.

A third variation of Poincar\'e patch coordinates is sometimes useful.  Let
$r = \log(u)$.  Then the metric becomes
\begin{equation}
ds^2_{AdS_{d+1}} = e^{2r} \left( -dt^2 + d\vec{x}^2 \right) + dr^2.
\end{equation}

\subsubsection{AdS-sliced coordinates}
First, parameterize $X_d = \arctan (\mu)$, where $\mu$ is an angle in
$[-\pi/2, \pi/2]$.   For the remaining hyperboloid coordinates choose any
coordinate system for an AdS space of one lower dimension with variable
radius given by $\cos \mu$.  The metric becomes:
\begin{equation}
ds^2_{AdS_{d+1}} = \frac{1}{\cos^2 \mu} \left(d\mu^2 + ds^2_{AdS_{d}} \right).
\end{equation}
We point out a useful identity from \cite{janus}.  If we take Poincar\'e
patch coordinates on the $AdS_4$ slices, the full metric is
\begin{equation}
\label{januspatchmetric}
ds^2_{AdS_5} = \frac{1}{y^2 \cos^2 \mu} ( - dt^2 + d \vec{x}^2 + dy^2 + y^2 d\mu^2),
\end{equation}
where $y$ is the warped coordinate on the $AdS_4$ slices and $\vec{x}$
refers to the two non-warped spatial directions on the $AdS_4$ slices. This
can be related to conventional Poincar\'e patch coordinates by the following
transformation:
\begin{align}
x &= y \sin \mu \\
\nonumber z &= y \cos \mu,
\end{align}
where $x$ is a non-warped direction in conventional Poincar\'e patch
coordinates.  For braneworld coordinates, this relation becomes
\begin{equation}
u = \frac{1}{y \cos \mu} = e^r
\end{equation}

For more details on the coordinate system see \cite{janus,fakesg}.  We will
use the term ``Janus-sliced coordinates'' as an easier to pronounce
alternative to ``AdS-sliced coordinates.''
\subsection{The Janus Ansatz}
The original Janus solution was an ansatz for a dilaton and 5-form running
in the warped direction of a deformed AdS, presented in AdS-sliced
coordinates.  The deformation alters the warp factor away from $1/\cos^2\mu$
and  introduces an ``angular excess'' \cite{janus}.

 When making the Janus ansatz, the warp factor of $AdS_5$ is promoted from the fixed function
$1/\cos^2 \mu$ to a more general $f(\mu)$.  The dilaton and 5-form are
allowed to run in the $\mu-$direction as follows:
\begin{align}
\phi & = \phi (\mu) \\
F_5 &= 2 f(\mu)^{5/2} d\mu \wedge \omega_{AdS_4} + 2 \omega_{S^5},
\end{align}
where the $\omega$'s denote unit volume forms on their respective subspaces.
With this ansatz, the supergravity equations of motion reduce to
\begin{equation}
\label{eq:dilaton}
\phi'(\mu) = \frac{c_0}{f^{3/2}(\mu)}
\end{equation}
for the dilaton, where $c_0$ is an integration constant, and
\begin{equation}
\label{januswarp}
(f')^2 = 4 f^3 -4 f^2 + \frac{c_0^2}{6 f}
\end{equation}
for the warp factor.  For more detail see \cite{janus,fakesg}.  Equation
\eqref{januswarp} can be integrated to find the maximum value of $\mu$,
dubbed $\mu_0$ \cite{janus}, and the integral can in turn be evaluated in a
series expansion for sufficiently small $c_0$ \cite{fakesg,janusdual}:
\begin{equation}
\mu_0 = \frac{\sqrt{\pi}}{2} \sum_{n=0}^\infty  \frac{\Gamma (4 n +
\frac{1}{2})}{\Gamma (3n +1) n!} \left( \frac{c_0^2}{24} \right)^n.
\end{equation}
It is worth noting that \eqref{januswarp} has 4 zeros in $f$, but only two
of them are real and only for the choice $0 \leq c_0 \leq \frac{9}{4
\sqrt{2}} \approx 1.59$.  So to find a physical solution, one must choose
the integration constant $c_0$ within that range and choose initial
conditions for $f$ so that it equals the largest root of the RHS of
\eqref{januswarp} at $\mu=0$.  The second face of Janus arises from
analytically continuing the solution to negative values of $\mu$.  This can
be implemented by brute force in numerics by taking the square root of
equation \eqref{januswarp} and multiplying the RHS by $\mathrm{sign}(\mu)$.

The main qualitative feature that distinguishes the Janus metric from
undeformed AdS is that the boundary occurs at a value of the warped
coordinate $\mu$ greater than $\pi/2,$ the so-called ``angular excess''
\cite{janus,fakesg,janusdual}.  The asymptotic behavior of the warp factor
near the boundary is $f(\mu) \approx \frac{1}{\sin^2 (\mu - \mu_0)} \left( 1
+ O(\mu-\mu_0)^{8} \right)$ \cite{fakesg}, identical at leading order to the
behavior AdS.  Non-perturbative stability of this solution was shown in
\cite{fakesg}.  The dual gauge theory was studied in \cite{janusdual}.  Many
subsequent papers have explored restoring supersymmetry in a Janus framework
\cite{superjan,D'Hoker:2006uu,D'Hoker:2006uv,
D'Hoker:2007xy,Suh:2011xc,Kim:2009wv,D'Hoker:2007xz},  adding a black hole \cite{janusbh2,janusbh},
nesting Janus spacetimes of different dimension \cite{janusinjanus}, and
many other variations \cite{Bachas:2011xa,Nishioka:2010ha,Chiodaroli:2010ur,Chiodaroli:2009xh,
Chiodaroli:2009yw,D'Hoker:2009my,D'Hoker:2009gg,Ryang:2009bs,Chen:2008tu,Honma:2008un,
Gaiotto:2008sd,Kim:2008dj,Azeyanagi:2007qj,Bak:2006nh,Sonner:2005sj}.

\section{Adding Probe D7 Branes with Janus-slicing}

Flavor is traditionally added to AdS/CFT via the prescription of adding
$N_f$ probe D7 branes that are spacetime filling in the AdS dimensions, wrap
a 3-cycle on the S5 and slip off the pole of the S5 at some finite value of
the warped coordinate, thus ``ending in thin air'' at that location
\cite{kk}.  In the dual theory this corresponds to adding a number $N_f$ of
massive  $\mathcal{N}=2$ hypermultiplets in the fundamental representation
of the gauge group.  The probe limit means that the number of colors, $N_c$,
is much greater than $N_f$, so any gravitational backreaction of the D7
branes on the metric can be neglected.  As noted in \cite{koy}, coordinate
systems other than Poincar\'e patch (or braneworld) do not readily admit an
interpretation as the near horizon limit of a black D3 brane (or stack of
$N_c$ coincident such branes). So we follow the example of \cite{koy} and
employ the strongest form of the AdS/CFT correspondence.  Even if the dual
gauge theory cannot be interpreted as the worldvolume theory of a stack of
coincident D3 branes, $AdS_5 \times S^5$ with D7 flavor branes in a
coordinate system other than Poincar\'e patch is a perfectly valid
supergravity system and should have a dual gauge theory description,
regardless of coordinate system or ansatz for the brane.  In \cite{koy}
flavor branes of various dimension were considered in global coordinates
with the ansatz that the branes could fluctuate only in the radial, or
$\rho$, direction.  We sill do the same in Janus-sliced coordinates, with
the ansatz that the branes can only fluctuate in the $\mu$ direction.  Much
heuristic intuition can be gained by considering results from flavor branes
in Poincar\'e patch coordinates and replacing $z \to \cos \mu$, effectively
fixing the $y-$coordinate to unity.  This can serve as a guidepost for certain
features, but many details are quite different in the Janus-sliced ansatz.
We proceed under the assumption that our description of the dual gauge
theory above is correct, exactly parallelling the original D3-D7 system
except the natural spacetime for constructing this dual gauge theory
consists of two copies of $AdS_4$ with their boundaries identified.  By
identifying boundaries, we mean that the boundary conditions for fields in
one $AdS_4$ are related to the boundary conditions in the other, similar to
the construction of \cite{connectivity}.

We apply the same procedure in Janus-sliced coordinates both with and
without the flowing dilaton of Janus itself.  The D7 brane extends in
the AdS directions and wraps an $S^3 \subset S^5$.  The D7 is only allowed
to fluctuate in the $\mu$ direction in Janus-sliced coordinates, and it
extends from the boundary to some non-zero value of $\mu$.  When the flowing dilaton
of Janus is added, we can no longer neglect the factor of the
dilaton in the DBI action, and the equation of motion picks up additional
terms from the dilaton.  Recall that $\phi$ denotes the dilaton field, not
an angle.  We use Poincar\'e patch coordinates on the $AdS_4$ slices with
$y$ as the warped coordinate of the $AdS_4$ slices and take $\psi$ and
$\theta$ to be the two angles on the S5 that are transverse to the D7 brane.
In regular AdS, the DBI action  for the Janus-sliced D7 brane has the same
form as the usual case, but the warp factor is slightly different:
\begin{equation}
\label{DBIreg}
S \sim \int d^8 x \,  \cos^3 \left( \psi(\mu)\right) \frac{1}{\cos^3(\mu)} \sqrt{1 + \cos^2(\mu) \left( \psi'(\mu) \right)^2}.
\end{equation}
The DBI action for full Janus can be obtained from \eqref{DBIreg} by
substituting $\cos \mu \to f^{-1/2}(\mu)$ and inserting the factor $e^{-\phi
(\mu)}$.

The resulting equation of motion is identical to the general equation of
motion  for D7 branes presented in \cite{holorg}, but we expand it here to
highlight unique features in Janus-sliced coordinates:
\begin{equation}
\label{eomreg}
0 =  3 \tan \psi(\mu) + \frac{ 3 \cos(\mu) \sin(\mu) \psi'(\mu)
+ 4 \cos^3(\mu) \sin(\mu) \psi'(\mu)^3 + \cos^2(\mu) \psi''(\mu)}{1 + \cos^2(\mu) \psi'(\mu)^2}
\end{equation}
Note that arcsine functions do not solve this equation of motion, unlike
the original flavor brane ansatz, as can be easily checked by plugging in
trial functions. We will turn to numerics to solve this equation of motion
in section \ref{regnumerics}.

\subsection{Asymptotic Expansion in Janus slicing}
\label{expansion}
To examine the near boundary behavior of $\psi$, we expand our equations of
motion about $\mu_0$. As in~\cite{holorg}, we will begin with the most
general possible form for a power series expansion of $\psi(\mu)$. We will
then substitute this into our EOM, expanded in terms of $v = \mu - \mu_0$,
and examine the restrictions the EOM place on the coefficients of the
expansion. According to the standard AdS/CFT dictionary, in order for the
field $\psi$ to properly describe a quark hypermultiplet mass operator in
the boundary theory, its near boundary behavior should follow,
\begin{equation}
\psi(\mu) \approx A (\mu - \mu_0)^{\Delta} + B (\mu - \mu_0)^{d - \Delta} ,
\end{equation}
with $\Delta = 3$. Thus, the leading behavior should be,
\begin{equation}
\psi(v) = Av + Bv^3.
\end{equation}

The general expansion for $\psi(v)$ has the form~\cite{holorg}:
\begin{equation} \label{eq:genex}
  \psi(\mu) = v \lbrace \sum^\infty_{n=0}\, \alpha_n v^n
                                   +\sum^\infty_{j=0}\,  \beta_j v^j \log(v)
                                   +\sum^\infty_{k=0} \sum^p_{l=2}\, \Psi_{k,l} v^k [\log(v)]^l \rbrace.
\end{equation}
We note that this expansion hinges on writing the metric in Fefferman-Graham
coordinates~\cite{holorg}, and in principle the coefficients of both the
expansion of the metric in Fefferman-Graham coordinates and the expansion of bulk fields can be functions of
other coordinates than the warped coordinate. The full implementation of
Fefferman-Graham coordinates in Janus-sliced coordinates was carried out in
\cite{skenderisjan}; however, for our brane ansatz, the bulk fields cannot
vary with any coordinate other than $\mu$, so we may safely ignore
dependence of the coefficients on AdS$_4$ coordinates. The Janus warp
factor, $f(\mu)$, has the same near boundary behavior as the $1/
\cos^2(\mu)$ of ordinary $AdS_5$, and may be treated as~\cite{janus,fakesg}
\begin{equation}
f(v) \approx \frac{1}{\sin^2(v)} \approx \frac{1}{(v-\frac{1}{6}v^3)^2} ,
\end{equation}
in the EOM. Furthermore, the dilaton field contributes to the EOM, only
through it's first $\mu$ derivative. Thus we can see from \eqref{eq:dilaton}
that it may be expanded as
\begin{equation}
\phi^{\prime}(v) \approx c_0\sin^3(v) \approx c_0 (v-\frac{1}{6}v^3)^3 .
\end{equation}
Below, we will keep the notation compact by denoting,
\begin{eqnarray}
X &=& \sin(v) \\ \nonumber
Y &=& -2\sin(v)\cos(v), \nonumber
\end{eqnarray}
and later expanding these trigonometric functions in their appropriate power series.

Our equation of motion for $\psi$ then takes on the near boundary form:
\begin{equation}
\label{eq:XYexp}
X^{2}\psi'' + [1+X^{2}(\psi')^2]3\tan(\psi) + \frac{3}{2}Y\psi' + 2YX^{2}\psi' - [1+X^{2}(\psi')^2]X^{5}c_0\psi' = 0.
\end{equation}
The explicit dilaton contribution is clearly identifiable, due to the factor
of $c_0$, and can be seen to contribute only at order $v^5$ and higher.
Since the leading asymptotic behavior of the warp factor is the same for
both the Janus and ordinary AdS solutions, and the explicit dilaton terms in
the Janus solution appear beyond the relevant order of $v^3$, we can
conclude that this expansion will proceed identically for both cases. Thus,
the asymptotic behavior found below, will apply to both the Janus solution,
and the "sliced branes'' in ordinary $AdS_5$.

In order to confirm the proper asymptotic behavior for $\psi$, we must
expand  the EOM to order $v^3$. When we explicitly expand~\eqref{eq:XYexp}
in powers of $v$, including all terms relevant up to order $v^3$, we find,
\begin{equation}
(v^2-\frac{1}{3}v^4)^2\psi'' + 3(\psi + \frac{1}{3}\psi^3) + 3v^2 \psi(\psi')^2 - (3v + 2v^3)\psi^{\prime}  = 0.
\end{equation}
At first glance, it seems there may be many contributing terms. However,
examining the EOM expansion~\eqref{eq:genex} at lower orders in $v$, and
looking at both the lowest, and the highest powers of $log(v)$, will reveal
some simplifying restrictions:
\begin{eqnarray}
0 &=& v \lbrace (-\beta_0 + 2\Psi_{0,2}) +(6\Psi_{0,3} - 4\Psi_{0,2} )\log(v) + \cdots \\ \nonumber
    &  & + (p-1)[p\Psi_{0,p} - 2\Psi_{0,p-1}][\log(v)]^{p-2} -2p\Psi_{0,p}[\log(v)]^{p-1} \rbrace.
\end{eqnarray}
Recall, that $l$ denotes an integer greater than $1$. If we assume that
$\Psi_{0,l}$ vanishes for $l = p$ and higher, then the second to last
$log(v)$ term tells us that $\Psi_{0,p-1}$ must also vanish, and so on down
the line. If there had been a contribution proportional to $ \beta_0 v
\log(v)$, this would have allowed a linear combination of $\beta_0$ and
$\Psi_{0,2}$ to vanish as the coefficient of the $v\log(v)$ term. However,
this is not the case, as the $\beta_0 v \log(v)$ terms from the portions of
the EOM linear in $\psi$ and $\psi^{\prime}$ cancel with one another. As
noted in~\cite{holorg}, an infinite tower of terms higher order in $\log(v)$
could exist, but this would invalidate the assumption of the existence of
the power series solution. The same argument can also be shown to apply to
the various $\Psi_{k,l}$ with $k \ge 1$. Thus, our power series solution for
$\psi$ actually has the simpler form,
\begin{equation}
\psi(v) = v \lbrace \sum^\infty_{n=0}\, \alpha_n v^n + \sum^\infty_{j=0}\,  \beta_j v^j \log(v) \rbrace.
\end{equation}
This mirrors the form found in~\cite{holorg}, with the radial  coordinate
$r$, which vanishes at the boundary, replaced by our $v = \mu-\mu_0$ which
also vanishes at the boundary ($\mu_0$ being equal to $\pi/2$ in the case of
ordinary $AdS_5$.)  Also note that notation for the expansion coefficients varies in the literature.  For example, \cite{holorg} uses $\phi_{(i)}$ where \cite{koy} used $\theta_{(i)}.$  We chose $\alpha_{(i)}$ to minimize confusion between expansion coefficients and angles corresponding to supergravity fields.

Inserting this simpler expansion for $\psi$ into the EOM, we can  examine
the leading order terms for further restrictions on the remaining
coefficients. The first and second order pieces give,
\begin{eqnarray}
v \lbrace -\beta_0 \rbrace &=& 0, \\
v^2 \lbrace -\alpha_1 - \beta_1 \log(v) \rbrace &=& 0.
\end{eqnarray}
From the lowest order term, it is clear that $\beta_0 = 0$. The  second
order contribution reveals that $\alpha_1 = \beta_1 = 0$. The vanishing of
$\beta_0$ ensures that there will be no higher powers of $\log(v)$
contributed by the $(\psi)^3$ and $\psi(\psi^{\prime})^2$ terms in the EOM.
Thus, the order $v^3$ contribution will be,
\begin{equation}
v^3 \lbrace \beta_2 -6\alpha_0 + 6\alpha_0^3 \rbrace = 0 .
\end{equation}
The coefficient $\alpha_0$ will thus be undetermined. The  coefficients
$\beta_0$, $\beta_1$ and $\alpha_1$ will vanish. Then, $\alpha_0$, when
fixed, will determine the value of $\beta_2$ though the algebraic equation,
\begin{equation}
\beta_2 = 6\alpha_0(1 - \alpha_0^2) .
\end{equation}
This same essential behavior (with a different algebraic equation relating
$\beta_2$ and $\alpha_0$) was found in \cite{holorg}. Insuring that it
applies here as well, confirms that our branes -- which are embedded
differently due to the radial coordinate $\mu$ picked out by the Janus
solution -- will still properly describe fermion flavors in the boundary
theory. This applies equally well in ordinary $AdS_5$ and Janus spaces.



\subsection{Numerics}
\label{regnumerics} 
\begin{figure}
\caption{\label{ubcombined} $\psi$ vs. $\mu$  (in red) and $\arcsin(\mu / \mu_b)$ (in blue), for $\mu_b = 0.5, 1.0, 1.3$}
\includegraphics{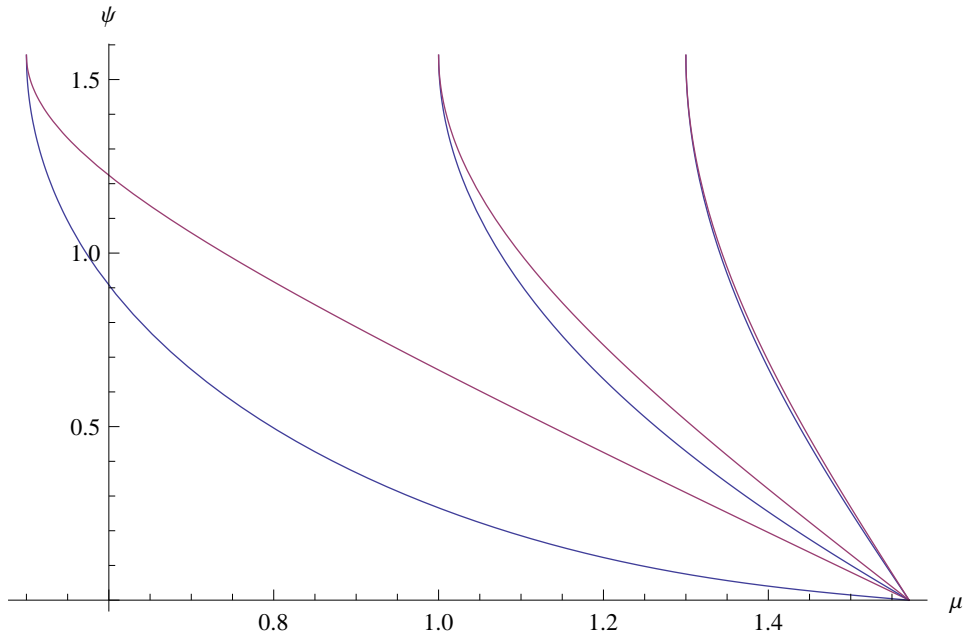}
\end{figure}

To solve \eqref{eomreg} numerically, we denote by
$\mu_b$ the ending location  of the brane and use as initial conditions
$\psi(\mu_b) = \pi/2 - \varepsilon,\ \psi'(\mu_b) = -1/ \varepsilon$ with
$\varepsilon = 10^{-3}$ to approximate the usual conditions that the brane
slip off the north pole with infinite derivative.  We present a few sample
solutions to the equation of motion in figure \ref{ubcombined}, the numeric solution is shown in
red,and the function $\arcsin \left( \cos \mu / \cos\mu_b \right)$ is shown in blue
for comparison.  We plot solutions for 3 different values of $\mu_b$ on the same axes, for reference.

For large $\mu_b$ (and thus large mass) the Arcsine function is close to the
actual solution, but it gets increasingly inaccurate as $\mu_b$ decreases.
Per the asymptotic expansion, we fit the numeric solutions to $\alpha_{(0)}
(\pi/2 - \mu ) + \alpha_{(2)} ( \pi/2 - \mu )^3$, where we have reversed the sign of the difference, $v$, introduced at the beginning of section \ref{expansion}.  The sign reversal is for convenience, so that we plot mostly positive masses, as an overall sign can be introduced in the dual gauge theory by a chiral rotation.  For reference, we list the
fit parameters for the example solutions shown in Figure \ref{ubcombined}.
For $\mu_b = 0.5$ we obtain $\alpha_{(0)} = 0.203805,\ \alpha_{(2)} = 1.21054$.
For $\mu_b= 1.0$ we find fit parameters of 1.40824 and 3.31012 respectively.
Finally for $\mu_b= 1.3$ we find 3.53835 and 12.2155.

\begin{figure}
\caption{\label{mvmu} Mass vs. $\mu_b$ for disconnected flavor branes}
\includegraphics{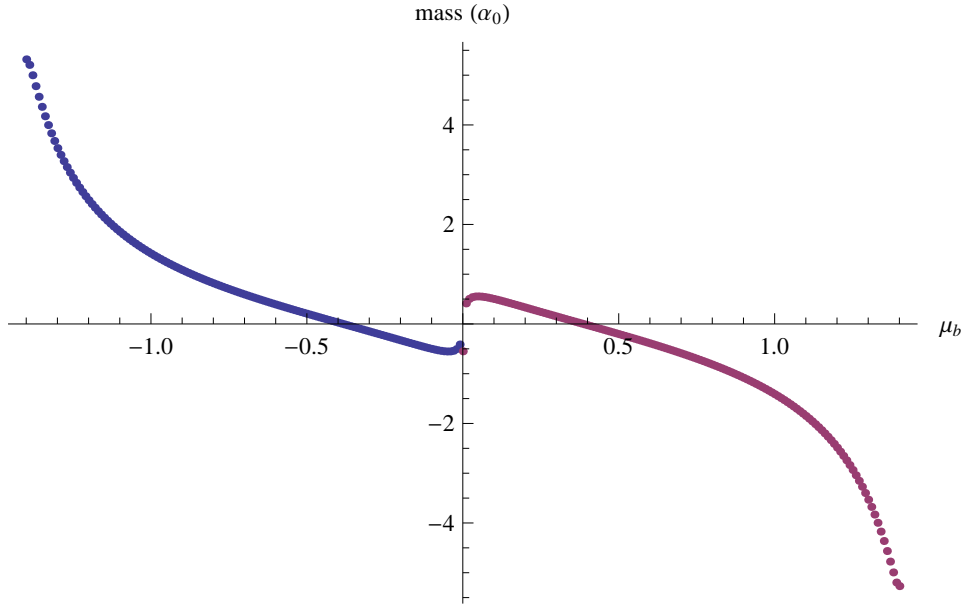}
\end{figure}

To obtain plots for studying the phase structure of this system, we
numerical solve and fit \eqref{eomreg} successively from $\mu_b = 1.4$ down
to $\mu_b=0.1$ in steps of 0.01, then adjust the step size  successively to
$10^{-5}$, $10^{-7}$, and $10^{-12}$ for reasons that will become apparent.
Recall that in \cite{kk} with arcsine solutions the mass was given by the
inverse of the position where the brane ended.  For Janus-sliced flavor
branes, the relation between mass $(\alpha_{(0)})$ and $\mu_b$ is more
complicated, given by Figure~\ref{mvmu}.
For values of $\mu_b$ near $\pi/2$, this has approximately  the same shape
as $1/\cos\mu_b$, as one might expect from naively extending the relation
from  \cite{kk} with Poincar\'e patch slicing using $z = y \cos \mu$, but
the relationship is more subtle for lower mass configurations.

\begin{figure}
\caption{\label{vevmasszoomout} $\alpha_{(2)}$ vs. $\alpha_{(0)}$ for disconnected flavor branes}
\includegraphics{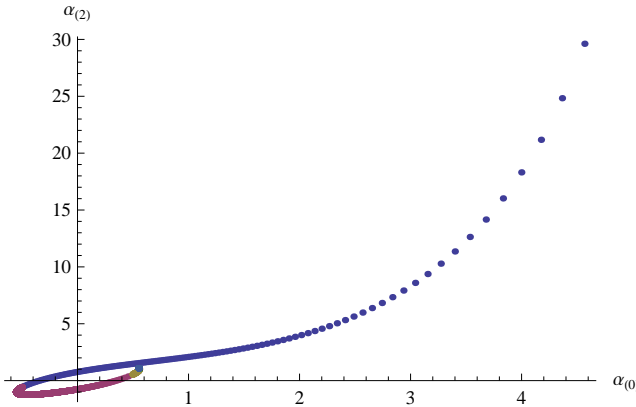}
\end{figure}

\begin{figure}
\caption{\label{vevmassplot} $\alpha_{(2)}$ vs. $\alpha_{(0)}$ for disconnected flavor branes detail}
\includegraphics{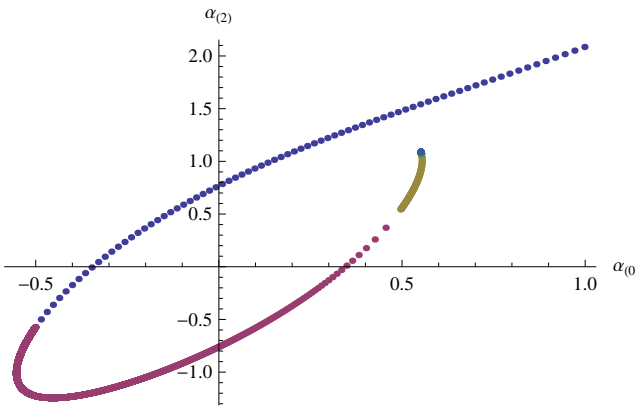}
\end{figure}

We plot $\alpha_{(2)}$ vs. $\alpha_{(0)}$ in Figure \ref{vevmasszoomout},
zooming in to give detail in Figure \ref{vevmassplot}.  The colors used
denote the step size taken in $\mu_b$.  The blue curve ends at $\mu_b
=0.01$; the red curve at $\mu_b=10^{-5}$; the yellow curve at $\mu_b=
10^{-7}$; and the green curve (which appears more as a dot) at $\mu_b=
10^{-12}$.  The step size for each color is equal to the ending value of
$\mu_b$.  At first, near $\mu_b =0.01$ and again near $\mu_b=10^{-5}$, it
appears that we are making bigger and bigger steps in the
$\alpha_{(2)}-\alpha_{(0)}$ plane as we approach $\mu_b=0$.  But then the curve
appears to stop abruptly at $\mu_b=10^{-12}$.

There appears to be a kind of limit point in the numerical solutions
approaching $\alpha_{(0)}= 0.5501, \alpha_{(2)}= 1.08895$ as $\mu_b \to 0$.
Note, however, that it is impossible to impose ``brane ending'' boundary
conditions at $\mu_b=0$.  The boundary condition that $\psi(\mu) \to \pi/2$
causes the $\tan \psi(\mu)$ term to diverge.  For non-zero $\mu_b$, the
accompanying boundary condition that $\psi'(\mu) \to \infty$ naturally
cancels this divergence when both are implemented as $\psi(\mu_b) = \pi/2 -
\varepsilon, \psi'(\mu_b) = 1/\varepsilon$.  Unfortunately, when $\mu_b =0$,
the coefficients of all the $\psi'$ terms in the equation of motion vanish,
reducing the equation of motion to $0 = 3 \tan\psi(\mu) + \psi''(\mu) /(1 +
\psi'(\mu)^2)$.  Imposing both boundary conditions simultaneously requires
infinite $\psi''(0)$ and cannot be solved numerically.  If we try a series
expansion near $\mu=0$, similar to the expansion in \cite{koy}, promoting
the infinitesimal $\varepsilon$ to a small fluctuation, $\psi = \pi/2 -
\varepsilon(\mu)$, the action reduces to
\begin{equation}
S ~ \int d^8 x \left(- \varepsilon(\mu)^3 \sqrt{1 + (\psi'(\mu))^2}\right).
\end{equation}
The resulting equation of motion is
\begin{equation}
\label{analytic}
\varepsilon^2 (\mu) ( -3 -3 (\varepsilon'(\mu))^2 + \varepsilon(\mu) \varepsilon''(\mu) ) = 0,
\end{equation}
which does not have an analytic solution except for the trivial one,
$\varepsilon=0$. The action and lack of a non-trivial solution are
consistent with the series expansions in \cite{koy} for branes of general
dimension that extend in $j$ of the AdS dimensions and $i$ of the $S^5$
dimensions.  Our action for $\mu \sim 0$ in Janus-sliced coordinates is
similar to the action of \cite{koy} for $i=2$, and the critical solution,
equation (3.4) of \cite{koy}, does not exist for the $i=2$ case.  This
analysis does not rule out the possibility of other solutions not detected
by our numerics that would continue the spiral of figure \ref{vevmassplot}
down to $\alpha_{(0)}=0, \alpha_{(2)} = 0$; however, if a non-trivial such
solution exists, it does not reach the origin of figure \ref{vevmassplot} at
$\mu_b=0$.  This would be extremely puzzling, since this is ordinary AdS
space where there is nothing to set a scale or otherwise mark any location
other than $\mu_b=0$ as special.

We find evidence of spontaneous chiral symmetry breaking in Figure
\ref{vevmassplot}.   The mass ($\alpha_{(0)}$) reaches zero between
$\mu_b=0.38$ and 0.39.   Since this occurs with $\alpha_{(2)}$ between 0.75398
and 0.784872, we conclude that the dual theory will exhibit spontaneous
chiral symmetry breaking as the mass of the quarks is taken zero. This would be expected for Janus proper, but it is surprising to find it in undeformed AdS simply with Janus-sliced flavor branes.  Interestingly, this chiral symmetry breaking is not detected by the test proposed in \cite{shock1,shock2}, but the geometry in our case is sufficiently different from the backgrounds considered there that this is not too surprising.  In particular, there is no central singularity in our geometry so the criteria used in \cite{shock1} do not apply.

\subsection{Asymmetric and Continuous Flavor Branes}
Janus-sliced coordinates do not exhibit a horizon, so the possibility exists
for a D7 brane to extend across the full range of $\mu$, from the
``right-hand'' boundary at $\mu=\pi/2$, through the ``center'' at $\mu=0$,
and out to the ``left-hand'' boundary at $\mu=-\pi/2$.  Indeed, the trivial
solution, $\psi(\mu) = \pi/2$, obviously exists and gives zero mass, zero
vev quarks in the dual theory\footnote{We are indebted to A.~Karch for these
observations.}.  Furthermore, the flavor branes examined in section
\ref{regnumerics} fill only half the spacetime, so quarks in the dual theory
would exist in only one of the two $AdS_4$ spaces.  Since the geometry is
symmetric under  $\mu \to -\mu$, it is trivial to extend these results by
adding a mirror image brane extending from $\mu= -\pi/2$ to $-\mu_b$.  This
is borne out by numerics:  numerically solving \eqref{eomreg} from
$\mu=-\pi/2$ to $\mu_b<0$ yields results for $\alpha_{(0)}$ and $\alpha_{(2)}$
that differ from the values for the corresponding positive $\mu_b$ by at
most $O(10^{-3})$.  For a given $\mu_b$ this would give quarks in the dual
theory with mass constant across both $AdS_4$ spaces.  However, the two
branes are  disconnected in the bulk, so there is nothing beyond aesthetics
imposing symmetry.  The D7 brane in the negative $\mu$ half of space may end
at a different distance than the D7 in the positive $\mu$ half of space.
The dual theory in such a case would have quark masses that were constant on
each $AdS_4$ but different between the two $AdS_4$ spaces.

\begin{figure}
\caption{\label{symsampplot} $\psi$ vs. $\mu$ for a continuous brane solution
with $\psi(0)=+1.4,\ \psi'(0)=0$.}
\includegraphics{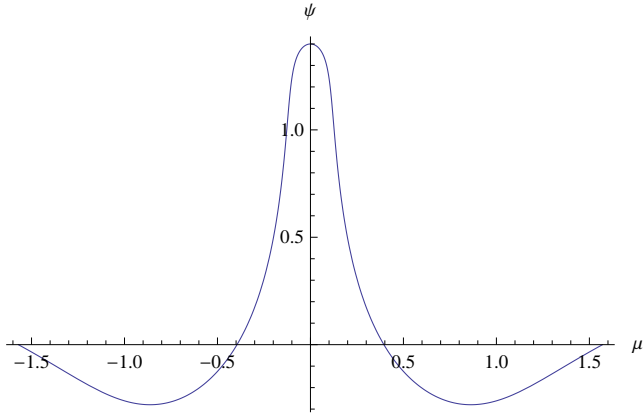}
\end{figure}

\begin{figure}
\caption{\label{ctssampplot} $\psi$ vs. $\mu$ for a continuous brane solution
with $\psi(0)=-1.4,\ \psi'(0)=+1$.}
\includegraphics{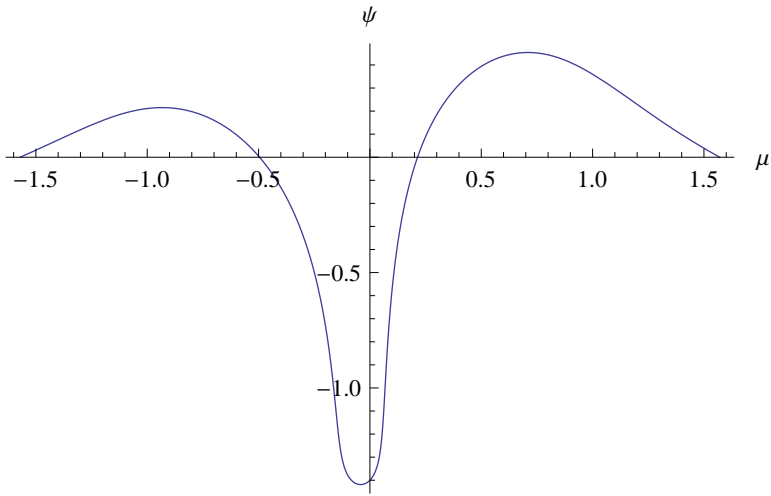}
\end{figure}

Continuous branes can also exhibit this asymmetry.  Generic solutions to
\eqref{eomreg} for continuous branes exhibit different asymptotic behavior
at the two boundaries.  If initial conditions are chosen for a continuous
brane solution such that the center of AdS, $\mu=0$, is a turning point for
$\psi$, then the symmetry of the geometry guarantees that the behavior of
$\psi$ will be symmetric at the two boundaries as well.  This, too, is borne
out by numerics to $O(10^{-3})$.  For continuous brane solutions, we solve
\eqref{eomreg} numerically imposing boundary conditions at $\mu=0$.  For
symmetric mass cases, we impose $\psi'(0)=0,\ \psi(0)=c$ where $c$ is a real
parameter we step through from -1.57 to +1.57.  For a sample solution in
this class, see figure \ref{symsampplot}.  For generic cases, we allow the
first derivative to be non-zero, giving a two-parameter family of solutions.
See figure \ref{ctssampplot} for a sample solution in this class.

\begin{figure}
\caption{\label{symmassvev} $\alpha_{(2)}$ vs. $\alpha_{(0)}$ for symmetric, continuous branes}
\includegraphics{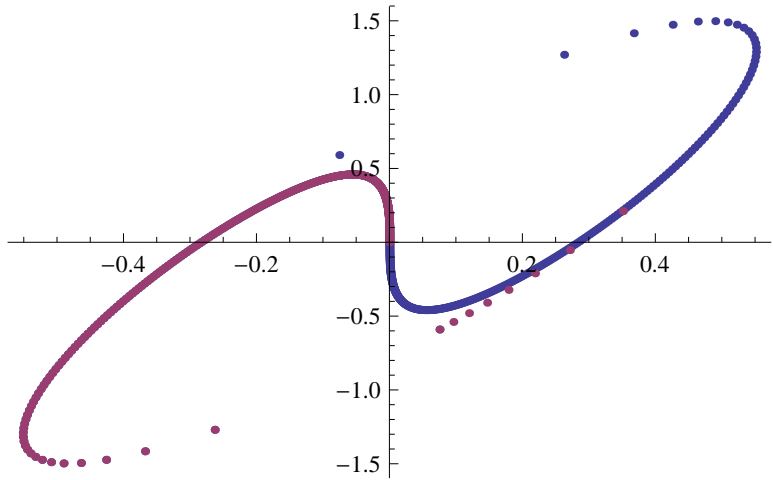}
\end{figure}

The easiest to understand data on continuous brane solutions come from the
symmetric case.  We plot $\alpha_{(2)}$ vs. $\alpha_{(0)}$ for symmetric
continuous branes as $\psi(0)$ ranges from -1.57 to +1.57 in steps of 0.005
in figure \ref{symmassvev}.  The plot begins at the origin for $\psi(0)=0$,
corresponding to the trivial solution. The two spiral arms are symmetric and
stem from the fact that changing the sign of $\psi(0)$ in this case simply
changes the sign of both fit parameters in the asymptotic solution.  The
blue curve represents solutions with $\psi(0) < 0$.  The red curve
represents $\psi(0) > 0$ and extends out to $\phi(0) = 1.5707$, with the
step size reducing to $\Delta \psi(0) = 10^{-4}$ after we reach $\psi(0) =
1.57$, then reducing again to $\Delta \psi(0) = 10^{-6}$ when we reach
$\psi(0)= 1.5707$. The fact that the red curve overlaps the blue so
completely is surprising. It certainly indicates that continuous branes
cannot produce quarks with arbitrarily large mass. The maximum value of mass
in our runs is $m=0.551489$. The near perfect overlap of the two curves
suggests that a phase transition may occur between different types of
symmetric, continuous branes: one with positive value of $\psi(0)$ very
close to $\pi/2$ and one with a more modest, negative value of $\psi(0)$
(and vice-versa).  The doubling back of the curves also indicates that
spontaneous chiral symmetry breaking occurs for continuous flavor branes
with $\psi(0) \lesssim 1.57$ where it appears the curve crosses the
$\alpha_{(2)}$ axis with non-zero intercept, indicating a non-zero vev at
zero mass.  Note that while this seems a value extremely close to $\pi/2$,
it occurs well before the region of the possible phase transition.

\begin{figure}
\caption{\label{contourleft} $\alpha_{(2)L}$ vs. $\alpha_{(0)L}$ and $\alpha_{(0)R}$ for general continuous branes}
 \includegraphics{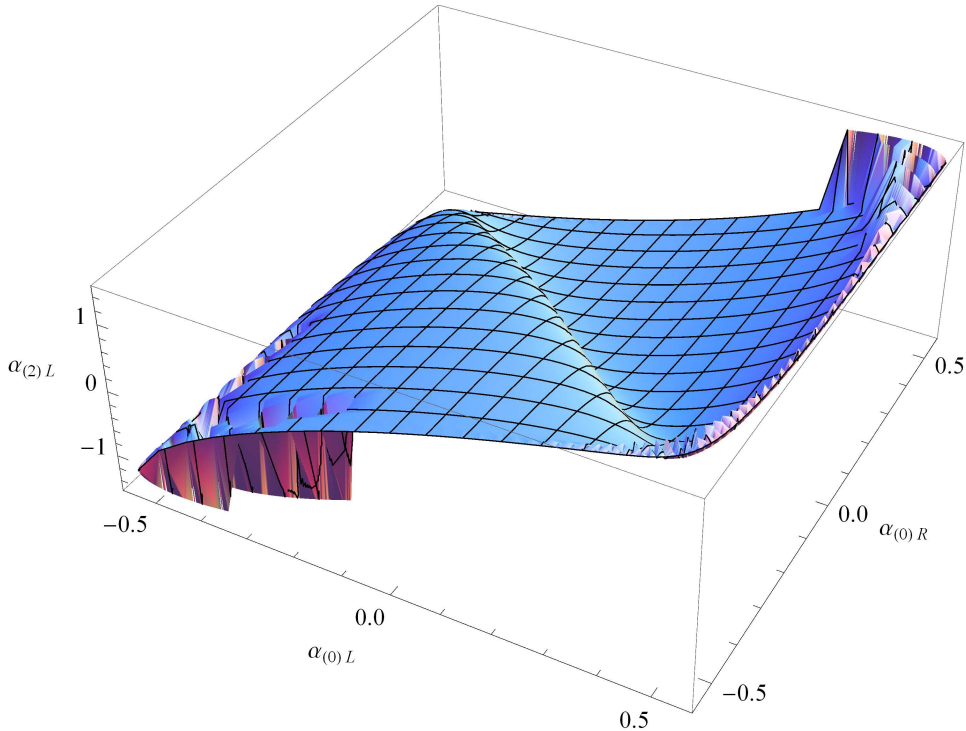}
\end{figure}

\begin{figure}
\caption{\label{contourright} $\alpha_{(2)R}$ vs. $\alpha_{(0)L}$ and $\alpha_{(0)R}$ for general continuous branes}
 \includegraphics{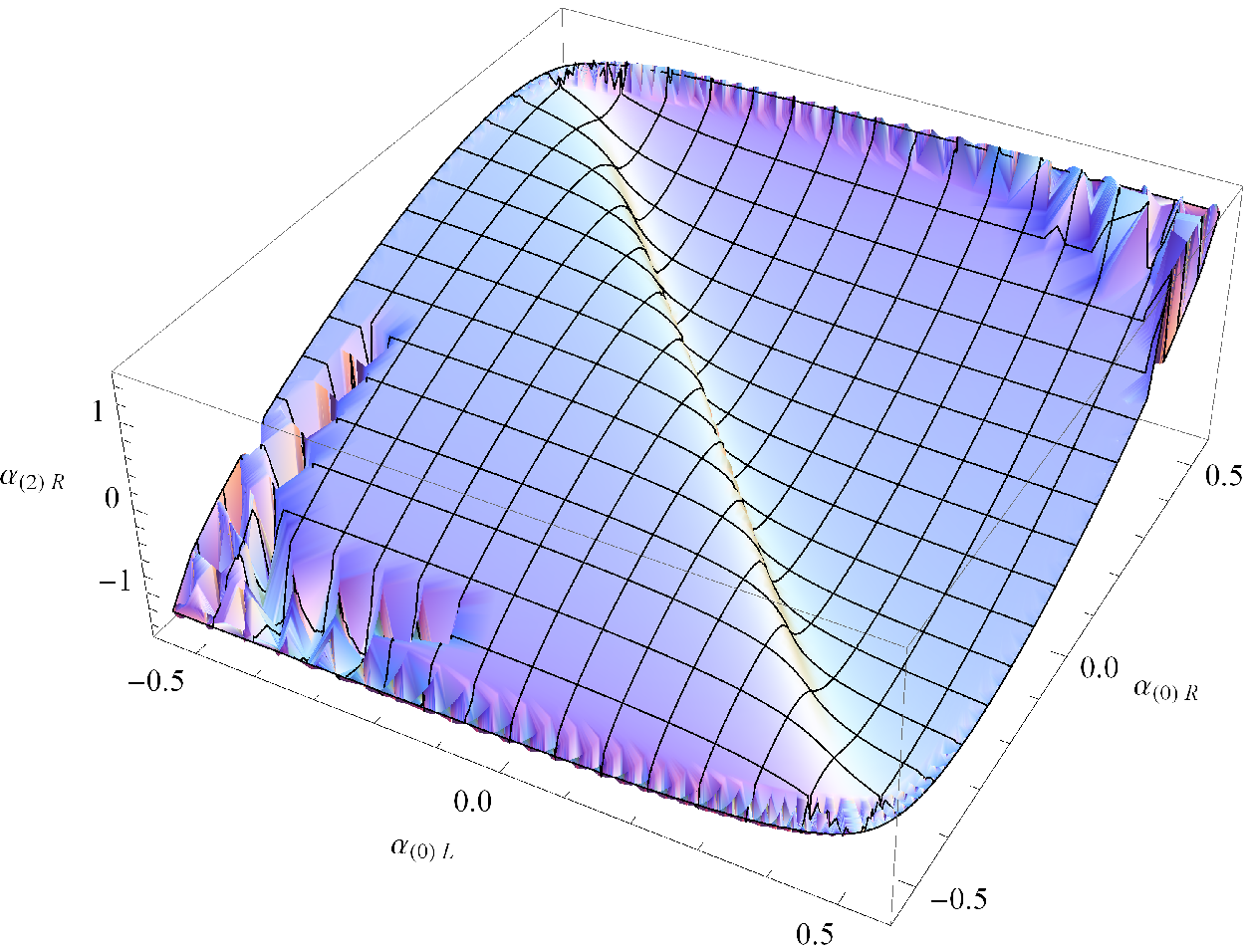}
\end{figure}

More generic continuous brane solutions require us to fit the asymptotic
solution at $\mu=+\pi/2$ and $\mu=-\pi/2$ separately.  We use subscripts
``R'' and ``L,'' respectively to denote these different sets of fit
parameters.  We present our results as a pair of contour plots, figures
\ref{contourleft} and \ref{contourright}, treating $\alpha_{(2)L}$ and
$\alpha_{(2)R}$ as dependent variables that depend on the pair of
independent variables $(\alpha_{(0)L},\alpha_{(0)R})$. Note the apparent
mirror image relationship between the two contour plots.

\section{On the Apparent Lack of Phase Transition between Disconnected and Continuous Flavor Branes}
\begin{figure}
\caption{\label{bothplot} $\alpha_{(2)}$ vs. $\alpha_{(0)}$ overlay plot for both  continuous and symmetric, disconnected branes}
\includegraphics{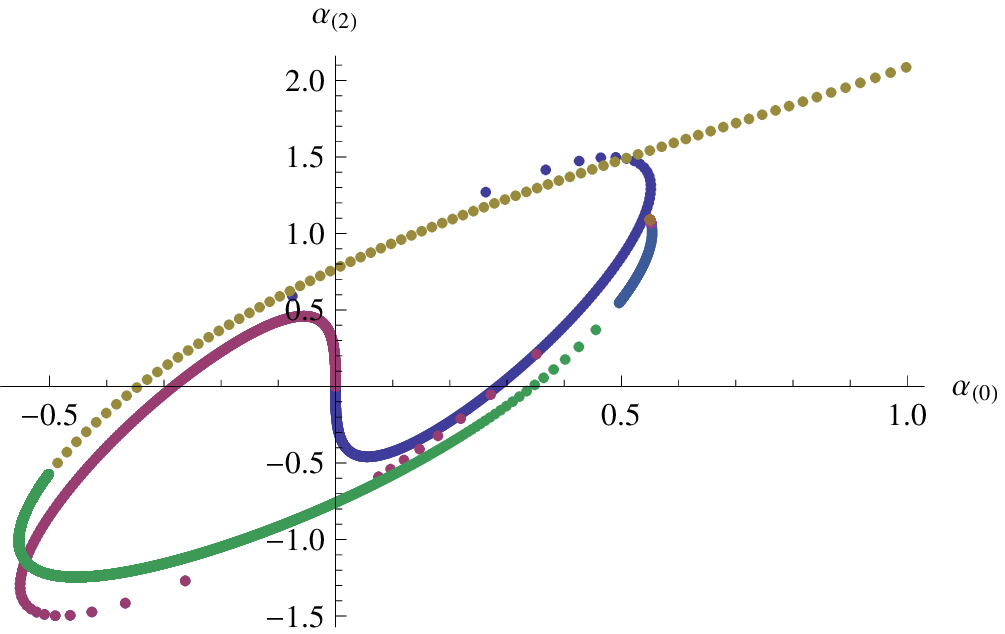}
\end{figure}

The natural expectation is that for low mass continuous branes might be
favored  while for larger mass disconnected branes might be favored.  This
does not prove to be the case, however.  In figure \ref{bothplot} we overlay
the plots from figures \ref{vevmassplot} and \ref{symmassvev}, but using
blue and red to denote the symmetric brane configurations.
While the curves cross at various points, we do not see the tell-tale
merging of  the plots that would signal a phase transition. We tried looking
for other classes of solutions to fill in the phase diagram, in particular
we considered disconnected branes that crossed the center of AdS at $\mu=0$
and ``bubble'' branes that pinched off at both a positive and a negative value
of $\mu$ and extend through the center.  Both seemed to be ruled out by
numerics.  Attempting to impose boundary conditions $\psi(\mu_b) \approx
\pi/2,\ \psi'(\mu_b) \approx \infty$ (necessary for both classes) while
solving for $0<\mu<\mu_b$ immediately encounters a singularity to the left
of $\mu_b$ when a numeric solution is attempted.

While it remains possible that there exists some class of solutions which we
have  not discovered, we speculate that the lack of phase transition is
because disconnected flavor branes and continuous flavor branes describe
incommensurate dual gauge theories. For example, with disconnected branes,
even if the masses are chosen to be equal, one need not even choose to use
the same number of branes on the left and right sides of AdS.  The fact that
this option is simply unavailable in the case of continuous branes is the
first piece of evidence to suggest the two types of solutions are described
by qualitatively different dual gauge theories. A second piece of evidence
is the lack of disconnected brane solutions pinching off at $\mu=0$.
Heuristically, we would have liked to think of the (non-existent) phase
transition as occurring when flavor branes on the left and right sides of
AdS approached the center and merged. However, in section \ref{regnumerics}
we showed that disconnected branes simply cannot ``pinch off'' at $\mu=0$.
This was shown both at large scale from the failure of equation
\eqref{eomreg} to accommodate the ``brane pinching off'' boundary conditions
at $\mu=0$ and at small scale (small values of $\mu$) by the failure of
equation \eqref{analytic} to admit non-trivial solutions in the vicinity of
$\mu=0$.  If disconnected branes can never reach or cross $\mu=0$, then
there is no hope of achieving a phase transition by melding disconnected
branes from the left and right halves of AdS in Janus-sliced coordinates.

Since the disconnected branes are confined to one half of the bulk AdS$_5$, we speculate
that the dual gauge theories for disconnected and continuous brane solutions
differ in the boundary conditions for the quark hypermultiplet fields at the
boundary between the two AdS$_4$ halves of the boundary theory.  A
disconnected brane on the right side of AdS$_5$ seems to naturally
correspond to a quark hypermultiplet restricted to the ``right'' AdS$_4$ in
the dual gauge theory with perfectly reflecting boundary
conditions at the boundary between the two AdS$_4$ half-spacees.  Whereas a continuous flavor brane
should correspond to quarks that can freely traverse from the ``right''
AdS$_4$ to the ``left,'' corresponding to ``transparent'' boundary
conditions where the leading and sub-leading terms on the two boundary
AdS$_4$ spaces match.  If this is the correct interpretation of the dual
gauge theory, then there clearly cannot be a phase transition as the two
scenarios are completely different.  However, this hypothesis poses
additional puzzles. If continuous flavor branes indeed describe an
$\mathcal{N}=2$ hypermultiplet with ``transparent'' boundary conditions
between the two AdS$_4$ spaces, there should be no obstacle to large quark
mass solutions.  Yet the continuous brane solutions do not seem to admit
solutions with quark mass larger than about $m_{L,R}=0.551$.  While it is
difficult to see this in the contour plots of figures \ref{contourleft} and
\ref{contourright}, it remains true for generic, asymmetric continuous
branes as well.  The most likely answer is that a second class of continuous
flavor brane solutions exists that was not detected by our numerics that
admits large mass solutions, although the space of possible
alternative explanations is by definition infinite.

\section{Flavor Branes in Janus}

The dynamical factor of the DBI action for such a D7 brane in Janus is
\begin{equation}
\label{DBIjanus}
S \sim \int d^8 x \, e^{-\phi (\mu)} \cos^3 \psi(\mu) f^2(\mu) \sqrt{f(\mu) + \left( \psi'(\mu) \right)^2}.
\end{equation}
This gives rise to the following equation of motion for $\psi$:
\begin{equation}
\label{janusquarkeom}
    \begin{array}{rl}
    0 & = \frac{f(\mu) \cos^2 \psi(\mu)}{2 \left( f(\mu) + (\psi'(\mu))^2
        \right)^{3/2}} \Biggl( 6 \sin \psi(\mu) f(\mu)^2 \left( f(\mu) + \psi'(\mu)^2 \right) +  \cos(\psi(\mu)) \Bigl(4 \psi'(\mu)^3 f'(\mu)  \\
        & + f(\mu) \left( 3 \psi'(\mu) f'(\mu)  - 2 \psi'(\mu)^3 \phi'(\mu) \right) + 2 f(\mu)^2 \left( - \psi'(\mu) \phi'(\mu) + \psi''(\mu) \right)  \Bigr) \Biggr).
    \end{array}
\end{equation}
Qualitatively, turning on the dilaton of Janus does exactly as little as we
expect it to thus far:  the warp factor is changed slightly, the boundary is
pushed out beyond $\mu=\pi/2$, and the equation of motion for our
Janus-sliced flavor branes picks up some extra terms proportional to the
derivative of the dilaton field.

\begin{figure}
\caption{\label{janusquarkplots} $\alpha_{(2)}$ vs $\alpha_{(0)}$ for Janus.
Green: $c_0 = 0.2$, blue: $c_0 = 0.8,$ red: $c_0= 1.2$.}
\includegraphics[totalheight=0.4\textheight]{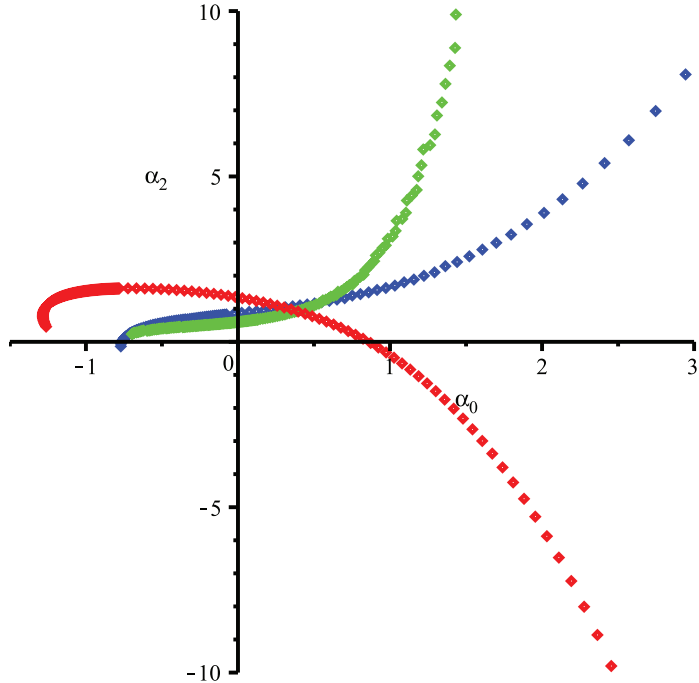}
\end{figure}

Numerically solving \eqref{janusquarkeom} with the same ``brane pinches
off'' boundary conditions as in regular AdS for $c_0 = 0.2,\ 0.8,\
\mathrm{and }\ 1.2$ we obtain figure \ref{janusquarkplots}.
The left end of each curve corresponds to $\mu_b= 0.2$.  Note that
the same qualitative shape found for Janus-sliced flavor branes in undeformed
AdS is present.  As $c_0$
increases, the shape of the curve changes more and more radically, changing
convexity and flipping over to the fourth quadrant for $c_0=1.2$.

\section{Implications in dual theory}
\label{sec:dual}
Roughly speaking, the natural choice of conformal factor in
AdS slicing gives a dual theory on two copies of $AdS_4$ joined at their
common boundary.  If we make this choice for Janus-sliced flavor branes in
undeformed AdS, we would expect to find that the quark mass jumps as one
crossed the boundary from one AdS$_4$ to the other.  After studying
Janus-sliced flavor branes in detail, we realize that far more dramatic
changes could take place, such as changing the number of flavors. Different
choices of conformal factor and thus different coordinate systems in the
dual gauge are of course allowed.

Following the general prescription of \cite{wittenholography} for
constructing the dual gauge theory and the refinements of
\cite{vijay1,vijay2}, we obtain the boundary metric by multiplying the AdS
metric by $g^2$, where $g$ is any function with a linear zero at the
boundary, often dubbed the ``conformal factor.''  We are specializing to the
case of Janus-sliced coordinates in the bulk. For a general bulk scalar
field, $\phi_\Delta$, of dimension $\Delta$, where $x$ generically denotes
the ``non-warped'' directions of AdS$_{d+1}$ has near boundary behavior
given by
\begin{equation}
\phi_\Delta \sim a_g(x) (g^{d-\Delta} + \ldots ) + b_g(x) (g^\Delta + \ldots).
\end{equation}
It is well known that $a_g$ is the source of the dual operator and
$b_g$ is the vev, although the vev is in general a more complicated function
of the leading and sub-leading coefficients using the prescription of holographic RG\cite{holorg}.  In the dual theory changing
metrics is accomplished with a conformal transformation.  In the bulk theory
the same is realized by choosing a different conformal factor to regulate the boundary metric.  To see the
relationships between coefficients with different choices of conformal
factor, recall in the bulk the field is a scalar, so each term in the
asymptotic expansion must be invariant.  If we compare two conformal
factors, $g_1$ and $g_2$, this gives us the relationship between two
boundary coefficients of
\begin{align}
a_{g1} &= \left(\frac{g_2}{g_1} \right)^{d - \Delta} a_{g2},\\
b_{g1} &= \left(\frac{g_2}{g_1} \right)^{\Delta} b_{g2}.
\end{align}
A similar discussion centered on the operator dual to the dilaton field
appears in \cite{janusdual}.  Applying this to Janus-sliced flavor branes
shows us that if we choose the conformal factor such that the dual theory
lives on two copies of AdS$_4$, then the mass of the quarks will be
piecewise constant.  If we conformally transform to Minkowski space,
examining metric \eqref{januspatchmetric}, we see that the mass becomes a
function of $y$ in the dual theory:
\begin{equation}
m_{\mathrm{M_4}} = \frac{m_{AdS_4}}{y},
\end{equation}
since the scalar field describing our D7 brane is of dimension 3.  Massive
quarks in actual Janus will be further complicated by the presence of the
operator dual to the dilaton \cite{janusdual}, but at leading order this
will not effect the position dependence of the mass operator in the dual
theory.  We postpone more detailed study of the dual theory for future work.

Since the phase diagrams of disconnected brane solutions and continuous
brane solutions do not exhibit behavior characteristic of a phase
transition, we hypothesize that the two types of brane solution are dual to
qualitatively different gauge theories.  We propose that continuous brane
solutions are dual to an $\mathcal{N}=2$ hypermultiplet mass operator that
has ``transparent'' boundary conditions at the shared boundary in AdS$_4$,
that is the leading and sub-leading terms should be the same in both
half-spaces.  Our proposal for the system dual to disconnected flavor branes
is that this dual mass operator exhibits totally reflecting boundary
conditions:  ``disconnected'' flavor branes on the right AdS$_4$ don't have
any impact on fields in the left AdS$_4$ and vice versa.  We think this is
the most likely case for several reasons.  First, from the gravity side, we
could choose to use a different number of flavor branes on each side of
AdS$_5$, giving a different number of flavors in the two half-spaces in the
dual theory.  Second, since the value of the dual operator is determined by
the leading order behavior of the gravity state as it approaches the
boundary, the coupling of the mass operator from a flavor brane in the right
half of AdS$_5$ is literally undefined in the left AdS$_4$ of the dual
theory.  A brane on the right ($\mu >0$) does not exist on the left ($\mu
<0$), so the asymptotic behavior of that state as it approaches $\mu= -\pi/2$
is undefined.  In the bulk, causality demands that, quite literally, the
left brane does not know what the right brane is doing. We see no way to
implement this in the dual theory without totally reflecting boundary
conditions for the dual quarks in each half-space AdS$_4$.  In the dual
theory for both flavor brane scenarios, we must impose totally transparent
boundary conditions on the gluons as those states are only sensitive to the
existence or type of flavor branes through interactions with the quark
states.

\section{conclusion}
We have proposed and analyzed Janus-sliced flavor branes as the appropriate
system for studying the addition of flavor to the Janus solution.
Janus-sliced coordinates on AdS$_5$ produces a theory most naturally dual to
$\mathcal{N}=4$ SYM on two copies of AdS$_4$ joined at a common boundary.
The main motivation for studying such the system is to look at Janus itself,
but our numeric studies have uncovered rich and surprising structure even in
undeformed AdS$_5$ without turning on the flowing dilaton of Janus.
Janus-sliced coordinates admit possibilities that don't exist in other
coordinate systems:  continuous flavor branes that extend from one boundary
through the center of AdS out to the other boundary have been studied in
global coordinates \cite{koy} but are not possible in Poincar\'e patch, and
asymmetric flavor branes that end at different positions on the two
``halves'' of AdS do not seem to be possible in any other coordinate system.

We have demonstrated that both disconnected and continuous branes exhibit
spontaneous chiral symmetry breaking, each possessing a state with non-zero
vev at zero mass.  Disconnected flavor branes can produce states of
arbitrarily large mass, while continuous branes (whether symmetric or not)
yield solutions with mass bounded by approximately $m_{max} = 0.551$ in
units of the inverse AdS radius.  This is one of many puzzles associated
with the continuous flavor brane solutions, as there seems to be no
compelling reason to restrict the mass of quarks in the dual gauge theory,
suggesting the existence of yet another class of flavor brane solutions.
There may also be a phase transition between continuous flavor branes with
values of $\psi(0)$ near $+\pi/2$ and those near $-\pi/2$, indicated by the
overlap in the phase diagram of figure \ref{symmassvev}. Analysis of the
free energies of the brane configurations will be necessary to confirm the
existence of such a phase transition.

Since there does not appear to be a phase transition between disconnected
flavor branes and connected, we propose that disconnected flavor branes are
dual to quark hypermultiplets with piecewise constant mass on AdS$_4$ with
totally reflecting boundary conditions at the boundary of each AdS$_4$
half-space, whereas continuous flavor branes are dual to quark
hypermultiplets with piecewise constant mass and totally transparent
boundary conditions.  In both cases, the entire gluon multiplet must have
totally transparent boundary conditions.  This proposal for differing quark
boundary conditions on AdS$_4$ is further supported by gravity-side
arguments such as causal disconnection of left and right branes as well as
the fact that the number of flavor branes could be chosen to be different on
each side.

We have also presented a phase diagram for quarks in Janus proper, using
disconnected branes with $\mu_b >0$ and three different values of the Janus
parameter, $c_0$.  The additional terms in the equation of motion arising
from the flowing dilaton make the numerics intrinsically less stable in
Janus proper than in undeformed AdS with Janus-sliced coordinates, so we
cannot offer as much detail.  Qualitatively, we see expected behavior, with
large mass when $\mu_b$ is near $\mu_0$, and solutions very similar to
undeformed AdS when $c_0$ is small.  Large $c_0$ begins to change
qualitative features of the phase diagram, pushing the ``turnaround point''
for the spiral further to the left and reversing the convexity of the curve
when $c_0$ is large enough.  It will be important for future work to address
the mysteries surrounding the continuous flavor brane solutions, determine
whether a phase transition exists for ``near polar'' continuous flavor
branes, and study the dual theories we propose in section \ref{sec:dual}. It
is very surprising that simply changing the flavor brane ansatz in this
manner results in such a radical change of behavior. It will also be very interesting to study the exact mechanism by which chiral symmetry breaking occurs, both in the dual gauge theory and in the supergravity.

\section*{Acknowledgements}
ABC and NC would like to thank J.~Shock and B.~Fadem for helpful
conversations.  ABC would like to especially thank A.~Karch for many patient
and detailed conversations about the peculiarities of flavor branes with the
Janus-sliced ansatz.  The work of NC was partially supported by the Raub
Fund.  The work of ABC was partially supported by a Faculty Summer Research
Grant from the Office of the Provost of Muhlenberg College.

\bibliography{janusbib}
\bibliographystyle{h-physrev}
\end{document}